\documentclass[12pt,preprint]{aastex}
\usepackage{lscape}
\usepackage{verbatim}
\usepackage{color}
\usepackage{ulem}

\newcommand{\ACBc}[1]{\textcolor{black}{ #1}}

\begin{document}
\title{The Heavy Element Composition of Disk Instability Planets Can Range From Sub- to Super-Nebular}
\author{Aaron C. Boley$^1$, Ravit Helled$^2$, and Matthew J.~Payne$^1$}
\affil{$^1$Department of Astronomy, University of Florida, 211 Bryant Space Science Center, P.O. Box 112055, Gainesville, FL, 32611-2055, USA} 
\affil{$^2$Department of Earth and Space Sciences, University of California, Los Angeles, CA 90095Ð1567, USA}

\date{}



\begin{abstract}

Transit surveys combined with Doppler data have revealed a class of gas giant planets that are massive and highly enriched in heavy elements (e.g., HD149026b, GJ436b, and HAT-P-20b).  It is  tempting to consider these planets as validation of core accretion plus gas capture because it is often assumed that disk instability planets should be of nebular composition. We show in this paper, to the contrary, that gas giants that form by disk instability can have a variety of heavy element compositions, ranging from sub- to super-nebular values. High levels of enrichment can be achieved through one or multiple mechanisms, including enrichment at birth, planetesimal capture, and differentiation plus tidal stripping.  As a result, the metallicity of an individual gas giant cannot be used to discriminate between gas giant formation modes.  
\end{abstract}

\subjectheadings{planetary systems: formation -- planetary systems: protoplanetary disks -- hydrodynamics }

\section{Introduction}

Core accretion plus gas capture (e.g., Pollack et al.~1996) and direct formation by disk instability (e.g., Cameron 1978; Boss 1997, 1998) are both viable formation models for gas giant planets (see D'Angelo et al.~2010 for a review).  \ACBc{Direct formation by disk instability} is posited to be limited to the envelope infall phase (e.g., Class 0-1 YSO) of disk evolution, with fragmentation preferentially occurring at large disk radii (e.g., Stamatellos et al.~2007; Clarke 2009; Rafikov 2009; Boley 2009; Vorobyov \& Basu 2010; see Boss 2006 for an alternative view), while core accretion planets are envisaged to form in an evolved disk after the oligarchic growth of cores has had time to proceed. The formation of giant planets by gravitational instabilities (GIs)  in the outer disk does not necessary preclude the core accretion mechanism in the inner nebula, and gas giant formation may proceed through both channels in some systems.  Hybrid scenarios, e.g., core formation in transient clumps (core assist plus gas capture; see Boley \& Durisen 2010, hereafter BD2010) and enhanced core growth in GI-induced pressure maxima (\ACBc{Haghighipour \& Boss 2003a,b}; Rice et al.~2004; Durisen et al.~2005), may also be possible and could form a non-negligible fraction of planets.  \ACBc{GIs may also activate at small radii in a disk due to the build up of mass in dead zones (Gammie 1996; Armitage et al.~2001; Zhu et al.~2010), but there remains disagreement between some research groups whether disk instability can form planets directly at $r\sim5$ AU (e.g., compare the results of Boley \& Durisen 2008 with those of Boss 2005).}  With multiple modes of formation available, it is desirable to understand what observational properties can be used for discriminating between different planet formation theories.

Several observational tests have been proposed to determine the relative importance of formation mechanisms.  Boley (2009) suggested that there could be two broad populations of gas giant planets, with disk instability planets peaking at large disk radii and core accretion at small radii. In addition, he suggested that the ratio of planets at large radii to the number of planets at small radii would increase with decreasing metallicity because the core accretion model is more sensitive to the fraction of solids in a disk than is the disk instability model.  However, this simple picture could be complicated by several effects.  First, there is no definitive demarcation radius for disk instability to operate, as the behavior of gravitational instabilities (GIs) is sensitive to the star-disk mass ratio (e.g, Nelson et al.~1998; \ACBc{Boss 2010)}, to thermodynamics (e.g, Pickett et al.~2003, Mayer et al.~2004), to radiative boundary conditions (e.g., Boley et al.~2007b; Mayer et al.~2007; Cai et al.~2010) to opacity (e.g., Johnson \& Gammie 2003; Cai et al.~2006; Cossins et al.~2010; Meru \& Bate 2010), and closely
related to opacity, the cooling rate (e.g, Tomley et al.~1991, 1994; Gammie 2001; Rice et al.~2005). \ACBc{To complicate matters further, some of the results from these studies may also be significantly affected by the choice of simulation resolution (e.g., Boss 2005; Nelson 2006; Meru \& Bate 2011a,b; Lodato \& Clarke 2011).}  Second, disk instability planets can migrate (inward and/or outward) from their in-situ formation radius through disk-clump and/or clump-clump interactions  (see, e.g., Stamatellos \& Whitworth 2009; Boley et al.~2010).  Using HR8799 (Marois et al. 2008, 2010) as an example, we do not know whether the four planets in the system formed by disk instability and moved inward, formed by core accretion and migrated outward, or formed through both channels, creating a compound system.  Despite these complexities, a trend in the ratio of wide orbit planets to total planets may still be observable. 

In a study determining the likely formation mechanism for the planets in HR8799, Dodson-Robinson et al.~(2009) suggested that,  if wide-orbit planets form by disk instability, the frequency of these planets should not trend with stellar age.  If there is a trend, scattering is the most likely explanation for planets on wide orbits, as these systems would represent a transient state. Strictly, even if most wide-orbit gas giants are produced by scattering, this does not indicate the formation mechanism since planets formed by disk instability could migrate inward and scatter back outward. 

Apart from trends based on the spatial position of planets, planetary composition might give clues regarding the relative frequency of formation mechanisms.  Transit surveys combined with Doppler data yield radii and mass measurements, allowing the mean density to be determined.  The bulk density, along with structure models, can be used to infer the mass of heavy elements (e.g., Guillot et al.~2006; Burrows et al.~2007).  
These measurements have revealed a class of gas giant planets that are massive and contain a high fraction of heavy elements  relative to their host stars.  Examples include HD149026b, GJ436b, and HAT-P-20b (Baraffe et al.~2008; Bakos et al.~2010; Leconte et al.~2010), where HAT-P-20b ($\sim 7$ M$_J$), in particular, could have a heavy element mass $\sim 300 M_{\oplus}$.  It is tempting to speculate that such a planet lends support to the core accretion mechanism.  However, giant planets with a large degree of enrichment could, instead, pose a challenge. For example, it is still not understood why Jupiter and Saturn are so enriched in heavy elements (see recent discussions in Li et al.~2010, as well as for possible enrichment scenarios).  In order to form large cores, nebular gas must first become depleted in heavy elements.  Runaway gas capture will accrete heavy element-poor gas, leaving a gaseous envelope that is nebular or sub-nebular in composition.  If runaway gas capture can begin at core masses less than 10 M$_{\oplus}$ ({Movshovitz} et al.~2010), the problem may be exacerbated. 
Extending this scenario to high-mass planets, a 7 $M_J$ planet  will
not have any obvious enrichment signature, and subsequent enrichment
of some form seems to be necessary in the core accretion model. \ACBc{Even if the
planet formed in a region of a disk that has a surface density
enhancement of solids, at the onset of rapid gas accretion, the
planetesimals will be cleared as the planet makes a gap.  Any advantage
that core accretion had in the enhancement is not obviously extended
beyond the formation of a critical core. } \ACBc{Suggestions for enrichment, at least in the context of HD 149026b, include collisions between gas giants or a steady supply of highly eccentric ($e>0.94$) planetesimals (see Ikoma et al.~2006 for details).}

Regardless of any difficulties that core accretion might have in producing planets like HAT-P-20b, if disk instability cannot reasonably produce planets with high levels of enrichment, then heavy element content could be used to constrain formation models.  In this paper, we find that disk instability planets can have non-nebular composition through a combination of one or more mechanisms. While we hope to find a way of discriminating between formation channels for a given planet, heavy element mass alone
{\it cannot} be used for this purpose.  The paper is organized as follows: In Section 2, we discuss three mechanisms that can lead to non-nebular compositions among disk instability planets.  Each enrichment model has its own subsection, wherein we discuss enrichment at birth, planetesimal accretion, and tidal stripping.   Section 3 is used to highlight how these mechanisms may operate alone or in combination to form highly enriched planets like HAT-P-20b.  The results are summarized in Section 4.

\section{Planetary Enrichment}

In this section, we discuss three channels for the heavy element enrichment of disk instability planets.  These mechanisms are enrichment at birth, planetesimal accretion, and tidal stripping.

\subsection{Enrichment At Birth}
Giant planets can be enriched with heavy elements as they form. When the solids in the disk are very small, they remain coupled and well-mixed with the gas.  For this trivial regime, 
we would expect  a clump born by disk instability to contain roughly $M_Z
\approx \frac{Z}{0.02} \frac{M_{\rm frag}}{1 M_J} 6.4 M_{\oplus}$ of solids \ACBc{(noted by Boss 1997)}, where $Z$ is the mass fraction of the high-Z material, and 0.02 is taken to be solar metallicity.   This estimate 
includes all elements heavier than helium, so it is strictly an upper bound on the high-Z material available for enrichment, as some volatiles may remain in a gaseous form. If, instead, the solids are large enough to decouple
from the gas before fragmentation, a range of gas giant composition becomes possible (BD2010).  Consider the following limiting cases: (1) If solids have sizes such that the stopping distance is roughly the width of a spiral arm, material will collect exactly where clump formation is expected, which  can lead to a fragment that has a super-stellar composition.  (2) If the solids are very large, they will not dissipate their energy in the spiral arms, possibly leading to offsets between the gaseous and planetesimal components.  If fragmentation does occur under these conditions, the solids may not follow the contraction of the clump, giving disk instability planets the chance to have, at least initially, substellar composition.  

\subsubsection{Efficient Aerodynamic Capturing of Solids in Spiral Arms Followed by Differentiation after Fragmentation}

To highlight outcomes with large initial enrichment, we show several snapshots of the simulation SIM1mu from BD2010. We also present a new simulation with twice the mass in large solids, which we call SIM1mu1.5Z, where the mass in grains, i.e., the opacity, is kept the same.  \ACBc{To recapitulate, the simulations are run using CHYMERA, which solves the equations of hydrodynamics on an a fixed, Eulerian, cylindrical grid (see Boley 2007 for details).  The equation of state accounts for the rotational states of molecular hydrogen (Boley et al.~2007a), where the ortho- and para-hydrogen states are assumed to be in equilibrium due to the long orbital timescales in these simulations.  The radiative cooling approximation presented in Boley (2009) is used, and particles are evolved as described in BD2010.}  SIM1mu \ACBc{is of a} 0.4 $M_{\odot}$ disk surrounding a 1.5 $M_{\odot}$ star.  The combined mass fraction of metals, silicates, and ices is set to 0.02, where half of this mass is placed into 10 cm-size solids and half into small grains, which gives the opacity for radiative heating and cooling.  The maximum solid size in the distribution of small grains is set to $1\mu$m. The size 10 cm was chosen to investigate the case of efficient concentration in spiral arms, as this size of solids will easily be captured aerodynamically by spiral arms in the outer disk.  {The 50:50 division between small grains and large solids is arbitrary, and a different distribution can lead to a different mass of solids that is concentrated into spiral arms. Nonetheless, the fractional enrichment, i.e., the large-solids-to-gas ratio in the spiral arms as compared with nebular composition, is not expected to be highly sensitive to the total mass in large solids. 
At very high masses of large solids, there may be some break in the efficiency of aerodynamic capturing, as seen in, e.g., the streaming instability (Johansen et al.~2009).  We will return this point in the section 3. The simulation SIMmu1.5Z explores the effects of a larger fraction of mass in rock-size solids on the capturing process for a fixed small-grain distribution. }

The large solids are represented by $10^5$ individual particles that are distributed proportionally to the gas, with the initial disk extending between roughly 50 and 300 AU.  The disk is given an irradiation profile of $T_{\rm irr}=T_0(r/{\rm AU})^{-1/2} + 10$ K, which sets the background temperature as a function of radius $r$ in the disk.  The initial model is created with $T_0=600$ K, but is then dropped to 130 K for the evolution.  This change transitions the disk from the regime that is expected be unstable to GIs , but self-regulating, to violently unstable with a strong chance of fragmentation.  \ACBc{The detailed Toomre $Q$ plots are given in BD2010, with a minimum $Q=1.5$ before the temperature change and a $Q<1$ before fragmentation}.  The setup was chosen to effect multiple clump formation in a single disk, and should be thought of as an experiment rather than representing exactly how disks are expected to fragment.  SIM2, which is discussed in the next subsection, uses initial conditions that are more in line with how we expect disks to be driven toward instability.  \ACBc{Although the change in the temperature profile may seem to be dramatic, we remind the reader that the spiral instability sets in when $Q\lesssim 1.7$ (e.g, Durisen et al.~2007) and fragmentation is not expected to occur until $Q<1.4$ (e.g., Nelson et al.~1998; Mayer et al.~2004).  Fragmentation is the result of a disk that is driven to a state of violent instability, which is what the change of the irradiation temperature is intended to represent. }
For SIM1mu1.5Z, we keep the number of particles the same, but increase their mass by a factor of two. The simulation begins on the high-resolution grid (see BD2010 for details), which corresponds to about 500 yr before the first panel in Figure 1.  The resolution in the vertical and \ACBc{radial} directions is 1 AU, and the azimuthal direction is divided into 512 zones. 

Figure 1 shows the formation of the first clump in SIM1mu  and its destruction 200 yr later. During the formation of the clump, 55 $M_{\oplus}$ of rock-size solids differentiated to the center of the fragment.  When the clump is tidally disrupted, the solids are released into the disk.  Because an amalgamation routine  is not included in the simulation\footnote{A subgrid model for amalgamation will be highly dependent on the assumptions we make for the outcome of collisions between particle ensembles.  It is not the purpose of this work to follow the evolution of masses and sizes of solids in these simulations, but to show how particles of a certain size behave.}, the formation of a bound, large core could not be followed. Figure 2 features three separate clump interactions, with the left panel corresponding to about 300 yr after the right panel in Figure 1.  The boxes A and C show mergers of multiple fragments, combining both gas and differentiated solid cores, while box B shows one clump that becomes destroyed after a glancing encounter with another.  The clump in box C grows to 32 M$_J$ with 270 M$_{\oplus}$ of total solids (including the well-mixed grains) as a result of the collisions, while the surviving clump in box B hovers around 11 M$_J$ with $\sim100$ M$_{\oplus}$ of total solids for the period shown. 

\begin{figure}
\includegraphics[width=7.5cm]{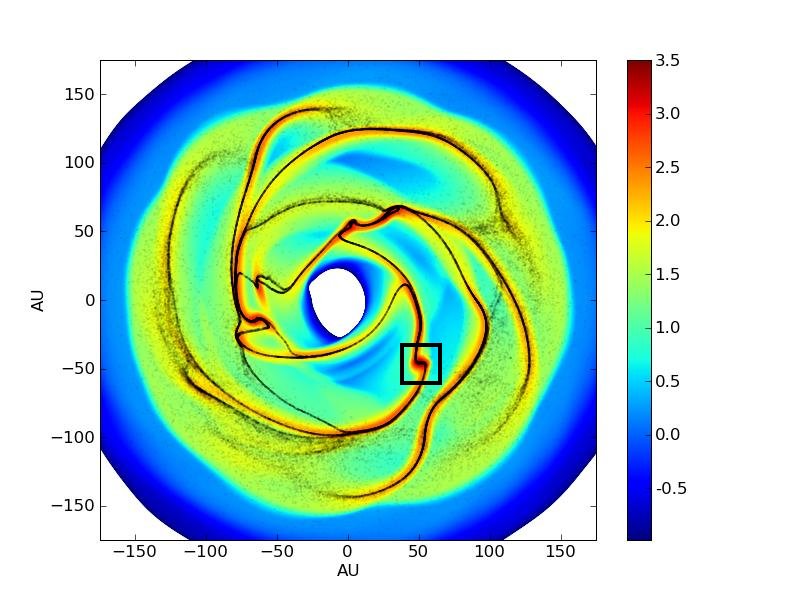}\includegraphics[width=7.5cm]{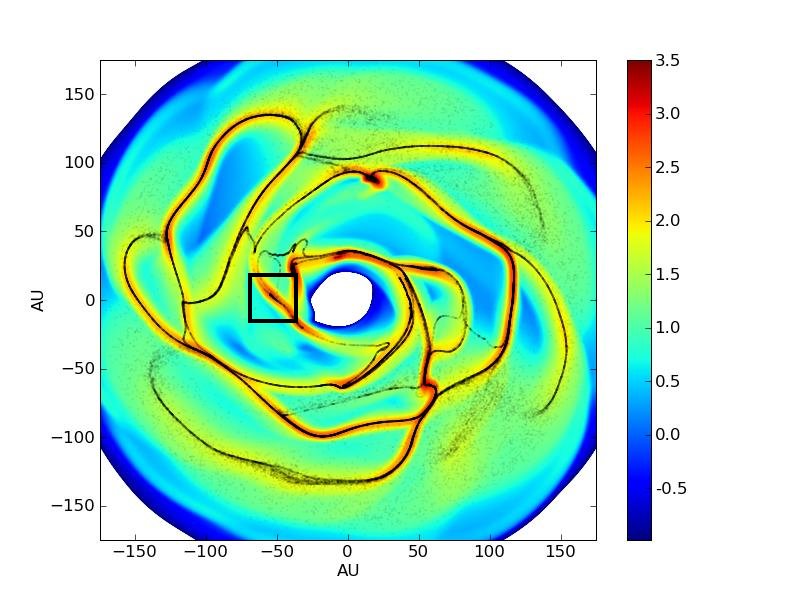}
\caption{Gas surface density (colorbar, log [g cm$^{-2}$]) with particle positions overplotted.  Left: A clump forms and solids are quickly marshaled to the center. The clump grows to mass of about 8 M$_J$ with 77 M$_{\oplus}$ of heavy elements.  Right: The clump is sheared away by the disk, releasing the solids back into the disk.  If the the particles were allowed to amalgamate, a core may have formed prior to disruption. The snapshots are $\sim 200$ yr apart. }
\end{figure}

\begin{figure}
\includegraphics[width=7.5cm]{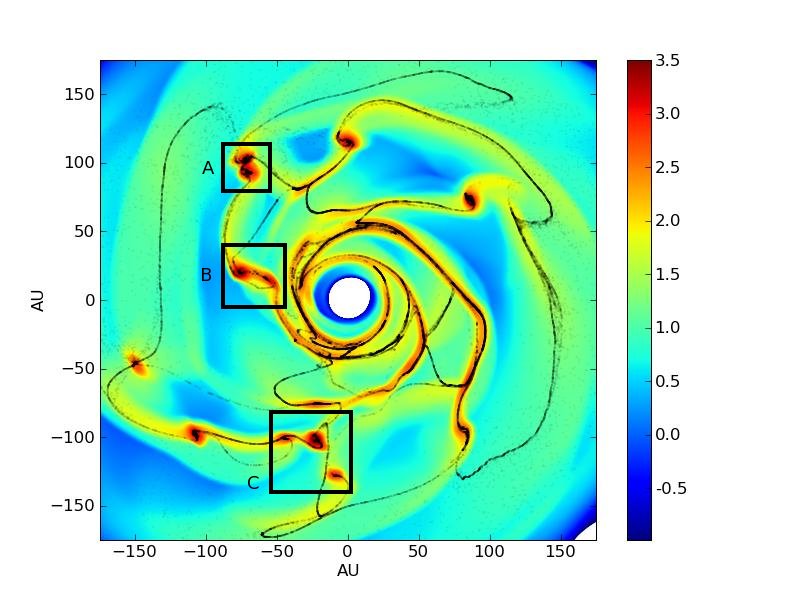}\includegraphics[width=7.5cm]{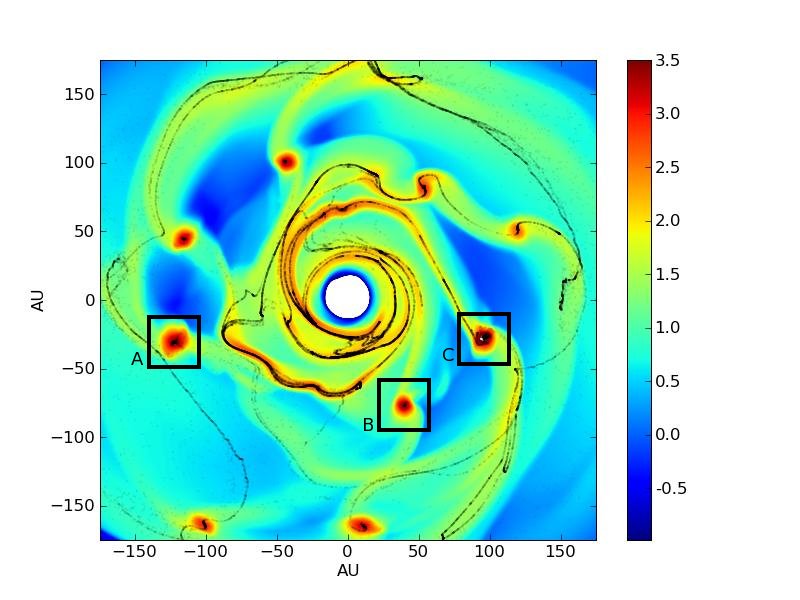}
\caption{Similar to Figure 1, but for later in the disk's evolution after additional fragmentation has occurred.  Left: There are three regions of interest that are highlighted by boxes A, B, and C.  Box A shows to clumps that are about to merge and become one object. Box B shows two clumps that just missed merging.  One is becoming disrupted, and releasing its gas solids back into the disk. In Box C, a three clumps are about to merge.  Right:  The boxes represent the same objects 200 yr later.  Boxes A  and C show that the clumps have completed their mergers.  The clump in Box C is now 32 M$_J$ with 270 M$_{\oplus}$ of total heavy elements. In contrast, one of the clumps in Box B has survived, while the other was tidally disrupted.  The clump is 11 M$_J$ with $\sim100$ M$_{\oplus}$ of heavy elements. }
\end{figure}

The evolution of simulation SIM1mu1.5Z is shown in Figure 3, using snapshots that are similar to those in Figure 2.  The increase in mass of large solids has led to differences between individual clumps, but the general behavior remains the same.  The solids are highly concentrated in the spiral arms, and when clumps form, they differentiate.  The clumps in Box A are very similar to those in Figure 2, but the clumps in Box B merge for SIM1mu1.5Z, instead of one destroying the other.  This particular clump in the 1.5Z simulation is also advanced in its orbit.  In Box C, left panel, one of the small clumps has already merged with the massive fragment.  At the end of the simulation (right panel), the clump in Box C is 26 $M_J$ with $\sim370$ $M_{\oplus}$ total in solids, giving the clump a total enrichment of about 1.5 over the nebular value.  Although the clumps have different final masses, the enrichment remains about the same for the large solids in the 1.5Z case.   The differences between the Box C clumps are summarized in Table 1. For the small region of parameter space that we have explored, the amount of solids captured by enrichment at birth seems to scale with the fraction of solids that can be aerodynamically captured by the spiral arms.  The result is consistent with the enrichment being pushed toward twice the nebular value in the limit that all solids can be aerodynamically captured by spiral arms.   

This capture efficiency may not continue to scale to very large metallicities, e.g., when the back reaction of the large solids on the gas becomes more appreciable in the dense, spiral arms.  
For example, one large difference between the SIM1mu and SIM1mu1.5Z is the frequency of knotty structure in remnant arms of solids, i.e., the solids that have been efficiently concentrated in a gaseous arm that has since dispersed.  When these arms enter regions where the gas density is very low, they can dominate the mass. This seems to cause the solid arms to form clumps, which is much more apparent in SIM1mu1.5Z than in SIM1mu.  The separation between adjacent knots can be many cells, so it is not obviously a grid-driven artifact.  However, the knot frequency might very well be related to the particle number that is used, as $10^5$ will lead to some shotnoise effects.  Even if the size scale of the clumping is a result of the particle resolution, the general behavior may be correct and related to the dependence of the streaming instability on metallicity (Johansen et al.~2009).  {If the metallicity between two disks is identical, then different distributions between small and large grains could have a similar effect.} We intend to explore this behavior in subsequent work, as this may lead to a break in the behavior of enrichment at birth for very high fractions of large solids relative to the gas. 

\begin{figure}
\includegraphics[width=7.5cm]{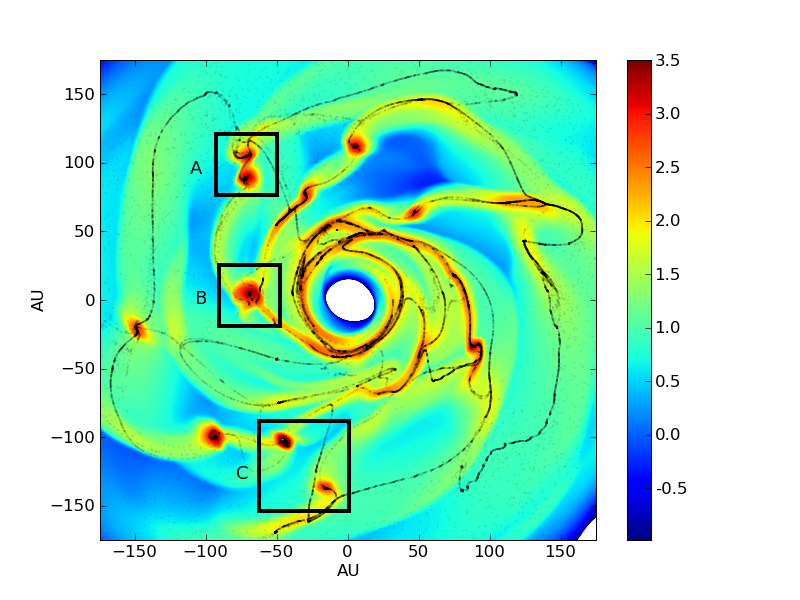}\includegraphics[width=7.5cm]{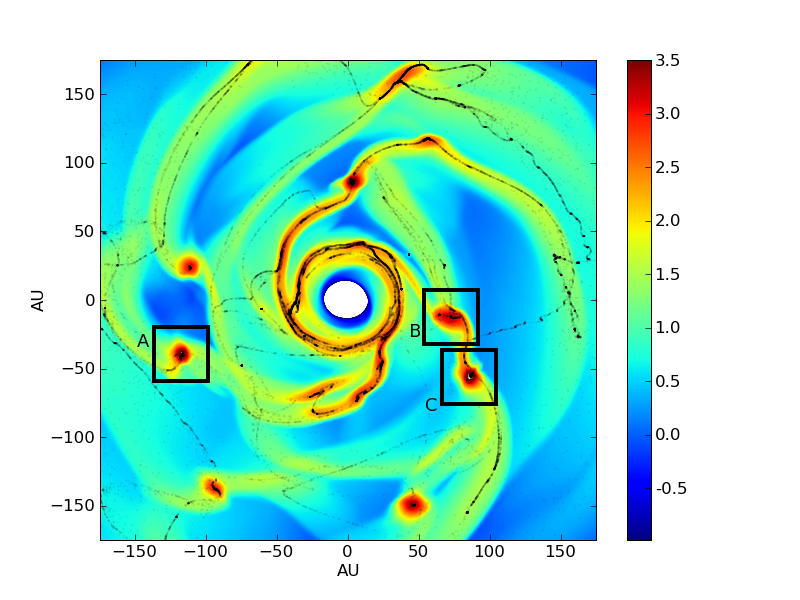}
\caption{Similar to Figure 2, but for SIM1mu1.5Z. The clumps in Box A are very similar to those in Figure 2, but  the clumps in Box B merge.  At the end of the simulation (right panel), the clump in Box C is 26 $M_J$ with $\sim370$ $M_{\oplus}$ total in solids, giving the clump a total enrichment of about 1.5 over the nebular value. }
\end{figure}

\subsubsection{Inefficient Aerodynamic Capturing of km-Size Planetesimals}

The simulations in the previous section explore the limit of efficient aerodynamic capture.  If, instead, the solids are decoupled from the gas, the initial enrichment can be much different. \ACBc{To demonstrate this effect, we choose to rerun} SIM2km of BD2010 with one modification: The initial circular speeds for large solids are determined based on the local radial potential gradient, instead of the Keplerian approximation used in BD2010.   Each particle is given a random vertical and radial velocity that is no more than 0.1 times the circular speed.  The disk model is a 0.33 $M_{\odot}$ disk surrounding a 1 M$_{\odot}$ star, and is a restart from SIMD in Boley (2009) about 1000 yr before fragmentation. The SIMD disk was driven toward instability by mass accretion from an envisaged envelope.  As the name suggests, the envisaged size of the particles is 1 km, so we call them planetesimals to distinguish them from the rock-size solids explored in the other simulaitons. We refer to this new simulation as SIM2kmv2. \ACBc{A comparison between the km-size results and the 10cm-size results for exactly the same initial conditions is presented in BD2010.  The simulations that are presented here are chosen to highlight limiting behavior succinctly.}

As found by BD2010, when all large solids are in km-size bodies, the particles do not concentrate in spiral arms and remain decoupled from the gas during the fragmentation process.  They can even form solid arms that become offset from the gaseous arms.  Table 2 shows clump properties for the first clump to form in SIM2kmv2, which is also the most massive.  For comparison, it also lists clump properties from SIM2km for one snapshot.  In both SIM2km and SIM2kmv2, the clump only has about $\sim 70$\% of the solids one would expect from a well-mixed nebula.   

While some planetesimals remain bound to the clump throughout its evolution, most of the km-size planetesimals that encounter the planet are not captured.  Figure 5 shows two snapshots of all planetesimals that ever pass through the clump.  Strictly, the number of planetesimals that we count is a lower limit because the analysis is post-process, with moderately coarse code outputs.  After encountering the planet, planetesimals are scattered inward and outward far from the encounter radius, showing a high degree of potential mixing.  Subtracting the total solids that were present in the clump for the first snapshot, we find that, during 1230 yr, 45 M$_{\oplus}$ of planetesimals pass through the extended clump, which has a volume-equivalent spherical radius of about 5 AU.  

\ACBc{The clumps in SIM2kmv2 produce a scattered disk through their interactions with
the planetesimals. Even if these clumps are eventually destroyed by tides, their
presence could play a large and possibly lasting role on the final distribution
of solids along the fringes of planetary systems.  We can take this result one
step further and speculate what type of objects might be produced by transient
clumps, particularly when they do not disperse simultaneously. If the first
clump differentiates and forms a core before its envelope is stripped, and the
core interacts with another clump, the core could be placed on a very
long-period orbit with a pericenter that is at the encounter semi-major axis.
In this case, a super-Earth-mass planet could be produced in outer planetary
systems with a pericenter $\sim100$ AU.}

\ACBc{If differentiation is incomplete, and only a rubble pile survives tidal
disruption, then, in principle, a swarm of dwarf-planet-size objects
could be produced.  Alternatively, planetoids could be formed directly
through aerodynamic trapping of rock-size solids in the spiral arms followed by
gravitational collapse of those solids (Rice et al.~2004,2006). The streaming
instability may also operate, as hinted by SIM1mu1.5Z. If planetoids can be
produced through any one of these mechanisms, then they too could
become scattered
by transient clumps and maintain large pericenters.}

\ACBc{The scattered bodies discussed here would likely have limited interactions with
the disk while the disk remains massive.  Using Sedna's orbital parameters as
an example, a scattered core/planetoid with a semi-major axis of
$\sim500$ AU will
have an orbital period of about 10 000 yr, assuming the mass inside pericenter
is about a solar mass. If the typical lifetime of a disk's
gravitationally unstable
phase is indeed $10^5$ yr, a Sedna-like scattered object will need to survive
ten pericenter passages as the
disk evolves. The survival of scattered cores/planetoids is also dependent on
whether the original scatterer has since become disrupted or has moved
from a given core's pericenter.}

\ACBc{The above discussion is not the focus of this paper, and follow up work must
certainly be conducted to determine the types of objects that can be produced
during disruption events followed by scattering during the early stages of disk
evolution. Keeping these issues in mind, we suggest that large cores, and
possibly planetoids, with both large semi-major axes and pericenters are
consistent with transient clump formation during the gravitationally unstable
period of disk evolution.}

\ACBc{Finally,} the simulations we present show that enrichment and even depletion at birth opens the possibility for disk instability gas giants to be born with a range of initial compositions relative to the nebular values, including super- and sub-nebular heavy element content.  The final composition will depend on the evolution of a clump and how it interacts with the disk.  Such interactions are the topics of the next subsections. 

\begin{figure}
\hspace*{-1.5cm}\includegraphics[width=7.5cm]{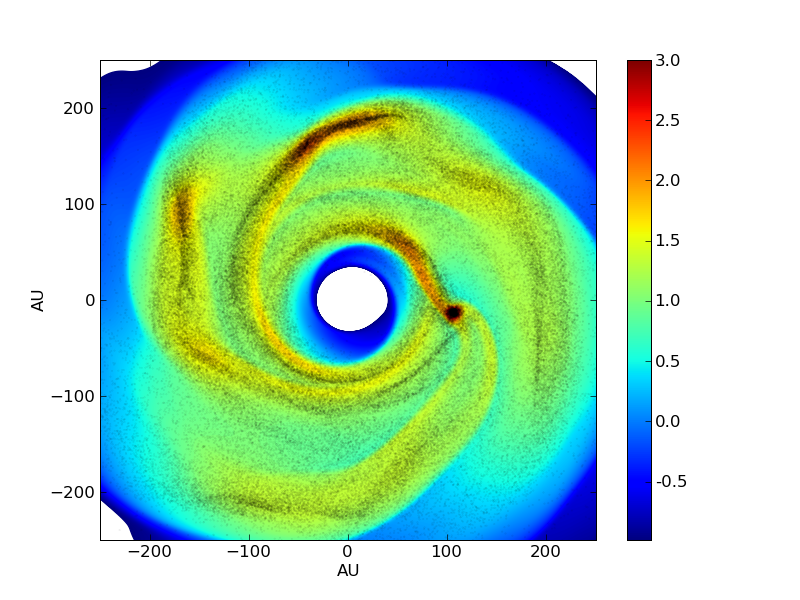}\hspace*{-1.25cm}\includegraphics[width=7.5cm]{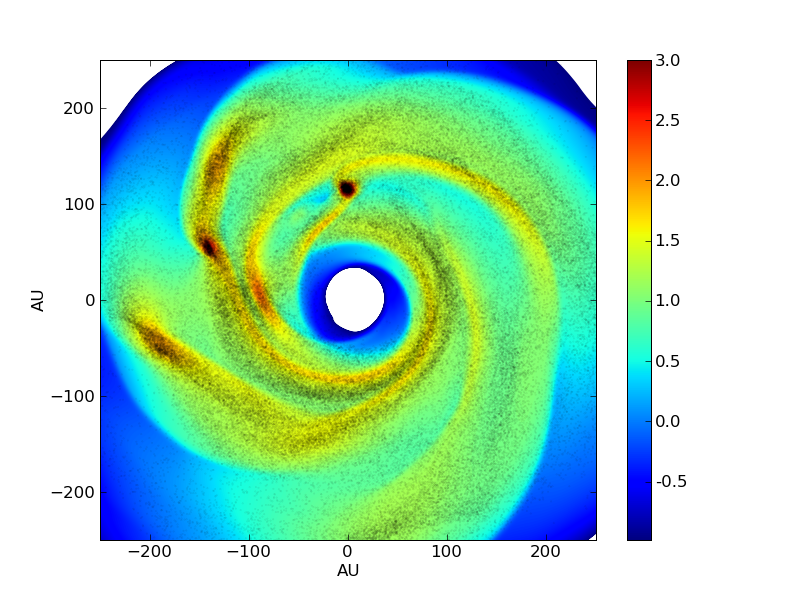}\hspace*{-1.25cm}\includegraphics[width=7.5cm]{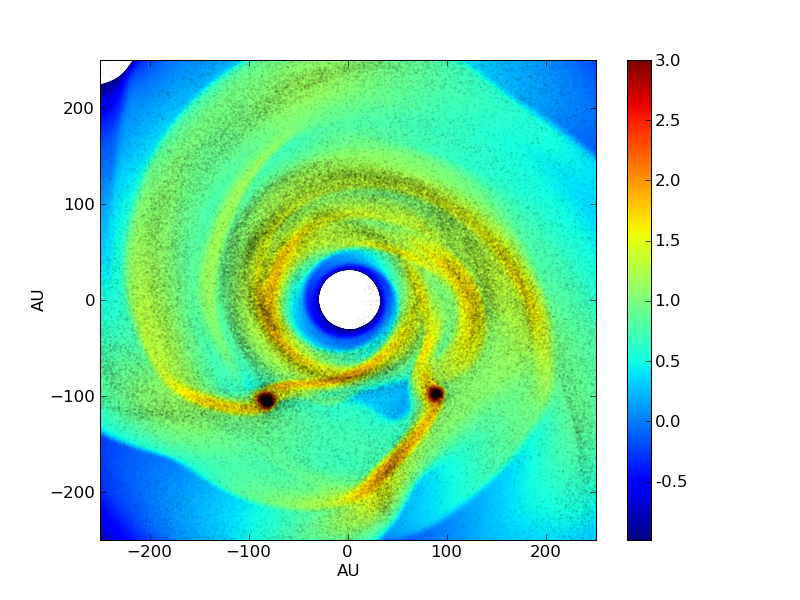}
\caption{Three shapshots of SIM2kmv2.  The center panel is roughly at the same time that SIM2km ends. As a result, BD2010 did not capture the formation of the second clump, but noted that the arm showed signs of fragmentation.  The solids do not collect in the spiral arms, and arms of planetesimals can even be offset from the gaseous arms. (Colorbar, log [g cm$^{-2}$])}
\end{figure}

\begin{table}
\begin{center}
\caption{Gas and large solid mass for the final clump in Box C in SIM1mu and SIM1mu1.5Z.  The disk radius $r$ and azimuthal angle $\phi$ of the clump are given for each data set, and can be compared with the right panels in Figures 2 and 3.  For SIM2kmv2, the clump positions correspond to the right and left panels of Figure 4.  As in BD2010, we consider mass to belong to the clump in regions where  the gas volume density $\rho>9\times 10^{-13}$ g cm$^{-3}$ and where the gas temperature $T>40$ K for SIM1 and $T>34$ K for SIM2.  The factor $f_{\rm StoG}$ gives the ratio of $M_{ls}$ to gas mass, where $M_{ls}$ is the mass in large solids (either 10 cm or planetesimals). The total enrichment of solids relative to the average nebular value is given by  $f_{\rm enr}\equiv(f_{\rm StoG}+0.01)/(0.01+0.01f_{ls})$, where $f_{ls}=1$ for SIM1mu and 2 for SIM1mu1.5Z.    In the last column, the notation $1.2(-10)$ refers to $1.2\times10^{-10}$. For our conversion to Earth mass, we use $1 M_J=320 M_{\oplus}$.}

\small
\begin{tabular}{ l l l l l l l l l }\hline
Simulation & Time & M$_{\rm gas}$ & M$_{\rm ls}$  & $f_{\rm StoG}$ & $f_{\rm enr}$ & $r$  & $\phi$ & $\rho_{\rm max}$ \\
& yr & M$_J$ &  $M_{\oplus}$ &  &  & AU & deg & g cm$^{-3}$\\ \hline\hline
SIM1mu  & 1870 & 32 & 170 & 0.017 & 1.3 & 100 & 344 & $1.2(-10)$\\
SIM1mu1.5Z & 1880 & 26 & 290 & 0.035 & 1.5 & 104 & 327 & $1.0(-10)$\\
SIM2km & 1430 & 7.1 & 7.2 & 0.0032 & 0.66 & 106 & 346 & $1.3(-11)$\\
SIM2kmv2&  1450 & 7.5 & 12 & 0.005 & 0.75 & 106 & 352 & $1.4(-11)$\\
SIM2kmv2 &  2340 & 13 & 18 & 0.004 & 0.70 & 132 & 232 & $2.3(-11)$\\\hline

\end{tabular}
\end{center}
\end{table}

\begin{figure}
\includegraphics[width=15cm]{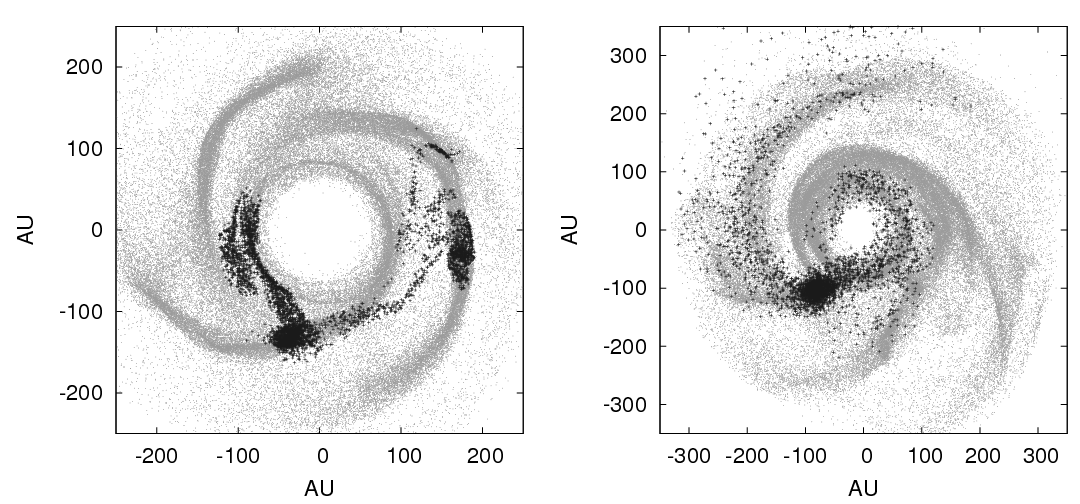}
\caption{Planetesimal distribution for two snapshots. Particles that encounter the planet at any time during the simulation are shown using black crosses. Left panel, 1110 yr: Planetesimals are accreted over a range of radii. Right panel, 2340 yr: After encountering the planet, the planetesimals have been scattered outward and inward, now ranging between $\sim$ 10s of AU to over 300 AU.}
\end{figure}

\subsection{Enrichment By Planetesimal Capture}

After fragmentation, the first evolutionary stage of the protoplanet (clump) is pre-collapse, i.e., the first core,
during which the extended ($\sim$ few AU) and cold clump contracts quasi-statically.  When the clump's central temperature reaches $\sim2000$ K, molecular hydrogen begins runaway dissociation, leading to rapid collapse of the clump down to a few Jupiter radii (e.g., Bodenheimer et al.~1980; Helled et al.~2006).  Using initial fragment properties from global hydrodynamics simulations,  the clump's first core stage can be followed in detail using a planetary evolution code, which solves the standard stellar structure equations. Although limited by simplifying assumptions, e.g., spherically symmetric, non-rotating, and isolated, such calculations can give insight on contraction timescales and observational signatures. They can also be used to estimate the size-scale of large solids that can be captured aerodynamically by the contracting structure. Helled et al.~(2006), Helled \& Schubert (2009), and Helled \& Bodenheimer (2010; hereafter, HB2010) investigated enrichment of clumps by planetesimal capture for a range of planetesimal sizes (1, 10, and 100 km) and disk locations (5-70 AU). The mass of solids that can be captured depends on the available heavy element mass in the clump's feeding zone, the cross section of the protoplanet as a function of time, the gas density of the clump,  and the sizes and velocities of the planetesimals.   

\subsubsection{Analytic Accretion Estimates Based On 1D Contraction Models}

The planetesimal accretion rate can be estimated by $dM/dt=\sigma \Sigma_Z \Omega F_g$ (Safronov 1972), where $\sigma$ is the protoplanet's capture cross section, $\Omega $ is the orbital frequency,  $\Sigma_Z$ is the surface density of solids, and $F_g$ is the gravitational focusing factor.   In the context of a gravitationally unstable disk, the surface density can be constrained by requiring that the Toomre (1964) parameter $Q= c_s\kappa/(\pi G\Sigma)\sim 1$ at the time of fragmentation at semi-major axis $a$. Here, $c_s$ is the sound speed, $\kappa$ is the epicyclic frequency, and $\Sigma$ is the gas density.  For simplicity, assume $\kappa\approx\Omega\approx\Omega_{\rm Keplerian}$ and the temperature profile of the disk $T=T_0 ({1\rm~AU}/a)^{q}$ K.  Using $Q$ and assuming a solid-to-gas ratio $f$, we can find a corresponding surface density.  The accretion rate can then be estimated by
\begin{eqnarray}
\frac{dM}{dt}&\approx &\sigma f  c_s \Omega^2 \left(\pi G\right)^{-1} \left(1+\frac{2GM_p}{\alpha^2v_{\rm Keplerian}^2 R_p}\right)\mathcal{F}_{3on2};\\\nonumber
& \approx &1.4\times10^{-4} \frac{f}{0.01} \left(\frac{\gamma}{5/3}\frac{T_0}{150\rm~K}\frac{2.3}{\mu}\right)^{1/2}\frac{M_{\star}}{M_{\odot}}\left(\frac{a}{1\rm~AU}\right)^{-3-q/2}\frac{\sigma}{\pi R_J^2}\\\nonumber
&\times&\left[1+\frac{4.3}{\alpha^2} \left(\frac{\sigma}{\pi R_J^2}\right)^{-1/2}\left(\frac{r}{\rm AU}\right)\left(\frac{M_p/M_{\star}}{10^{-3}}\right)\right]\mathcal{F}_{3on2}M_{\oplus}~\rm yr^{-1}.
\end{eqnarray}
The term in square brackets represents standard two-body gravitational focusing, where $\alpha$ is a parameterization for the planetesimal velocity dispersion, assuming the dispersion $\left< v\right>\approx \alpha v_{\rm Keplerian}$.  Protoplanet and stellar masses are $M_p$ and $M_{\star}$, respectively, and the ratio of the three-body to the two-body gravitational focusing enhancements is given by $\mathcal{F}_{3on2}$.
This ratio has been studied extensively by Greenzweig \& Lissauer (1990, 1992), but for regimes quite different than explored here. For the moment, we center our discussion around the case for which $\mathcal{F}_{3on2}$ is unity.   

%
%
 
 First, consider the post-collapse stage of the clump, where its radius is between 1 and 3 $R_J$.  Assume that $\alpha\sim0.1$, an assumption we will justify later this section.
Using equation (1), with all other values nominal, the timescale at 100 AU for a Jupiter-mass  planet to double its solids content from the well-mixed value of about  6 $M_{\oplus}$  would be about 3 and 1 Myr for $\sigma=\pi R_J^2\ {\rm and}~9 \pi R_J^2$, respectively. For a 10 $M_J$ planet, the timescales to double its solids by accreting 60 $M_{\oplus}$ of material is also 3 and 1 Myr,  {\it ceteris paribus}.  These timescales do show that planetesimal accretion can be important after the clump has collapsed. However, our estimated timescales are similar to the evolution timescale of the disk, which suggests a self-consistent study is required to address this phase properly.  We will return to this point in Section 2.2.2. 

In contrast to the post-collapse stage, the entire pre-collapse evolution is short enough to take place while the disk is gravitationally unstable, even with the large uncertainty in collapse times. For example, immediately after disk fragmentation, a clump's radius can be several AU in size.  Over a period of $10^3$ to $10^5$ yr, the clump will contract to a few tenths of an AU.  At this point, its center becomes hot enough to effect rapid collapse by H$_2$ dissociation (Bodenheimer et al.~1980; Helled et al~2006, 2008; Helled \& Bodenheimer 2010, 2011).  Contraction will be faster as the clump becomes more massive or metal poor, due to either the initial metallicity or due to grain sedimentation.   Planetesimal capture is not guaranteed during this phase because the densities in large regions of the protoplanet can be very low.   For this reason, Helled et al.~(2006), Helled \& Schubert (2009), and HB2010 used a planetary evolution code to follow the contraction of the first core and determine $R_{\rm cap}$ as a function of time, the aerodynamic capture radius for a given planetesimal size.  In Figure (6) we show the evolution of the $R_{\rm cap}$ for a 3, 5, 7, and 10 M$_J$ fragment.  The symbols represent data points taken directly from the HB2010 simulations. The simulations assume that the clumps are in isolation and have a well-mixed, solar composition. The data are for the case of 1 km-size planetesimals, where the relative speed of the planetesimals far from the clump are set to 0.1 times the Keplerian speed at 68 AU, the distance they were envisaged to be orbiting (see HB2010 for further details). Helled \& Bodenheimer (2011) extended their previous contraction calculations to include the effects of grain growth and sedimentation in the atmospheres. They found that grain sedimentation lowers the photosphere to significantly higher temperatures, shortening the contraction times for a range of masses to about 1000 yr.  {To model this effect, we also scale each collapse sequence to a time of 1000 yr}. 

\begin{figure}
\includegraphics[width=7.5cm, angle=-90]{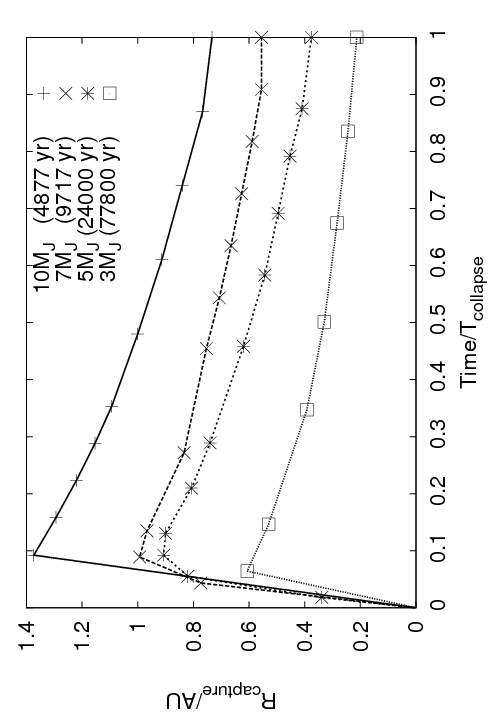}
\caption{Data (symbols) from the  HB2010 simulations of a 3, 5 , 7, and 10 $M_J$ protoplanet in isolation.  Each radius represents  each protoplanet's $R_{\rm cap}$ for 1 km planetesimals.   The evolution is normalized by the time of the last data output, which is roughly the time until dissociative collapse. Each $T_{\rm collapse}$ is given in parentheses on the plot, consistent with the HB2010 data. We also scale each collapse sequence to a time of 1000 yr to model the effects of sedimentation (Helled \& Bodenheimer 2011). At early times, the capture radius is negligible because the densities are small.  Eventually the capture radius becomes comparable to the size of the protoplanet, and follows the protoplanet's contraction. When the planetesimals are larger, they are captured less efficiently, and when smaller, more efficiently.}
\end{figure}

The HB2010 capture radii can be used  in equation (1) to explore the enrichment of disk instability protoplanets during the first core phase.  We set $M_{\rm star}=1.5 M_{\odot}$ and $T_0=350$ K, which gives a surface density of solids $\sim 1$ g cm$^{-2}$ at 68 AU, consistent with the value used by HB2010. While computing the accretion mass, the solid surface density is reduced by the accreted mass, assuming the mass is spread over an annulus with a width of 8 Hill radii and centered on the planet's location.  HB2010 only explore enrichment when $F_g=1$, but we complement the calculations to include  $F_g>1$, where $\alpha=0.1$ and 0.3.  The results are given in Table 3. 
 While checking for consistency between our results for $F_g=1$ and those of HB2010, we discovered a error in the HB2010 calculation that led to smaller values than we find here. Regardless, their conclusion that planets would have negligible enrichment by planetesimal capture at 68 AU remains valid, {\it if the effects of focusing are negligible.}

\begin{table}
\begin{center}
\caption{Mass capture results for 1 km-size planetesimals during the first-core phase using the same parameters studied by HB2010 ($F_g=1$; cf.~results by HB2010 -- see text for details).   We also include mass capture estimates including two-body gravitational focusing, using velocity dispersion $\alpha$s 0.1 and 0.3 times the Keplerian speed at 68 AU.  We define $f_{\rm cap}\equiv M_{\rm cap}/(0.02 M_{\rm planet })$, i.e., the mass captured relative to what is expected for the well-mixed case.  The column with $f_{\rm cap}(t_{1000})$ is for the scaled contraction time of 1000 yr.  Only $\alpha=0.1$ is shown because planetesimal capture is about 25\% of the well-mixed distribution by $\alpha=0.3$ for the fast contraction.  The accretion luminosity, which is only shown for the standard contraction case, is the planetesimal accretion luminosity just before collapse.  The peak luminosities for both dispersions is a factor of a few larger than the end state. For this table, we take $M_{J}=320 M_{\oplus}$.  }
\small
\begin{tabular}{ ccccccccc }\hline
M$_{\rm planet}$  &   M$_{\rm cap}$ &  M$_{\rm cap}$ & M$_{\rm cap}$  & $f_{\rm cap}$ & 
$f_{\rm cap}$ & $f_{\rm cap}(t_{1000})$& $L_{\rm acc}$ &  $L_{\rm acc}$ \\
M$_J$ & M$_{\oplus}$ & M$_{\oplus}$ & M$_{\oplus}$ & & & & $L_{\odot}$ & $L_{\odot}$\\
 & $F_g=1$ &   $\alpha=0.3$ & $\alpha=0.1$ & $\alpha=0.3$& $\alpha=0.1$& $\alpha=0.1$ & $\alpha=0.3$ & $\alpha=0.1$   \\\hline
3   & 16 &  130 & 580 & 6.8 & 30 & 0.76 & $6\times10^{-6}$& $10^{-5}$ \\
5  & 14 &  120 & 600 & 3.8& 19 & 1.3  & $2\times10^{-5}$& $5\times10^{-5}$\\
7  & 8 &  80  & 490 &  1.8 & 11 & 1.5 & $4\times10^{-5}$& $2\times10^{-4}$ \\
10 & 7 &  77 &  500 & 1.2 & 7.8 & 2.0 & $7\times10^{-5}$ & $4\times10^{-4}$ \\\hline
\end{tabular}
\end{center}
\end{table}

The focusing term itself is fairly modest (typically around ten for $\alpha=0.3$), but including the term makes the difference between a negligible and a detectable amount of heavy element enrichment for a range of conditions. 
This is emphasized by the results in Table 3.  When the contraction times are consistent with those of a well-mixed clump, even a high velocity dispersion of $\alpha=0.3$ can boost the solids content of a clump by factor of a few.  For $\alpha=0.1$, the mass accretion of solids can be well over a Jupiter mass.  When the contraction times are very short, velocity dispersions with $\alpha\sim 0.1$ still lead to planetesimal accretion masses comparable to twice the heavy element mass of a well-mixed clump.  
  Also shown in Table 3 are the estimated luminosities due to planetesimal capture for the standard contraction, where we take $L_{\rm acc}\approx GM_{\rm planet} (dM_Z/dt) / R_{\rm cap}$.  When $\alpha=0.1$, the accretion luminosity of solids is comparable to the luminosity of the clump in isolation (see HB2010 Fig.~1).
 Based on these estimates, we expect that the solids will affect the contraction of the clump by contributing a non-negligible amount of mass, by depositing energy in the envelope, and by potentially changing the opacity of the system, but addressing these points in detail is beyond the scope of this paper.  The main point of the above exercise is to demonstrate that, if there is a large population of comet-like planetesimals, disk instability planets can become moderately to significantly enriched during the first-core phase, even at large disk radii. 
 
 \subsubsection{Accretion Estimates Based on N-Body Simulations of  a Planet and Planetesimals in a Smooth Disk}

Greenzweig \& Lissauer (1990, 1992) ran a series of simulations to ascertain differences between the three and two-body gravitational focusing factors, represented here by $\mathcal{F}_{3on2}$.  They found that two-body focusing leads to a fairly accurate estimate of the actual focusing for large velocity dispersions, which is the case for a gravitationally unstable disk (next subsection).  Whether their results are valid for these relatively high-mass ratio protoplanets is unclear.  Moreover, without planetesimal orbital damping or mixing mechanisms, massive clumps can affect the surface density of solids, possibly limiting enrichment to only tens of orbits. 

We address these concerns by using Mercury (Chambers 1999) to integrate $10^5$ test particles in an annulus centered on a 10 M$_J$ gas giant at 100 AU.   The width of the annulus extends from 25 to 175 AU (10 Hill radii). We consider three sizes for the planet: 1, 3, and 418 (0.2 AU) $R_J$, corresponding, roughly, to a collapsed, cold gas giant, a gas giant immediately after dissociative collapse\footnote{The models of  Helled \& Bodenheimer (2011) find a planet size of about 2.4 $R_J$ after hydrostatic equilibrium is re-established following dissociative collapse in a 10 M$_J$ clump.}, and a gas giant just before dissociative collapse.  For the latter, the value underestimates the capture radius of a 10 M$_J$ clump (see Fig.~6).  Whenever a particle passes  within the planet's radius, the particle is considered to be captured and it is removed from the simulation.  The particles are given a Raleigh distribution for eccentricity, with the typical $e=0.05$.  The inclination $i$ is treated similarly, with the typical $i=e/2$.  The spatial distribution corresponds to a surface density that falls as $a^{-1.75}$, which is expected for a gravitationally unstable disk with near-Keplerian rotation and a temperature profile that drops as $a^{-0.5}$. The orbital phases are mixed. The total mass of planetesimals within the annulus is about 3 M$_J$, consistent with a gravitationally unstable disk with the nominal values used in equation (1).  All orbits are Keplerian.  The point of this preliminary study is to use a mass reservoir consistent with the context of disk instability, but then to consider only Keplerian orbits, ignoring the smooth and fluctuating components of the disk's potential.  In this way, we can compare the derived accretion rates with equation (1), and test for the possible importance of the nonaxisymmetric structure, as suggested by the full hydro+nbody simulations presented above.  

In Figure 7 (left) we show the surface density profile of planetesimals for one of the simulations as a function of time.  As expected, after 10 orbits, a gap is cleared, with the surface density reduced by about a factor of ten  immediately surrounding the planet.  After 100 orbits, the surface density is reduced by about 100.  The effect of this clearing on the accretion rate is shown in Figure 7 (right).  For the extended radius (418 $R_J$), the clump accretes at roughly the analytic rate for the first 2000 yr, for a reasonable range of $\alpha$.  After this time, the effects of gap clearing alter the accretion rate, and the accretion rate becomes much lower than expected.  Using the collapse times from HB2010, we expect a 10 M$_J$ clump to undergo dissociative collapse after about 5000 yr, so the variation in the surface density will have an effect on the total mass accreted.  For longer collapse times, this variation could become very significant. 

The self-limiting accretion is expected to be the same for the 1 or 3 $R_J$ planets because no change has been made to their potential relative to the 418 $R_J$ protoplanet.  Unfortunately, the initial mass accretion rate is too affected by shotnoise to allow comparisons to be made with the analytics for times less than about $10^4$ yr, as one particle is about 0.1 M$_{\oplus}$.  After this time, the cumulative mass is still subject to discreteness effects, but enough mass has been accreted to make rough comparisons with the expected values.  As with the 418 $R_J$ protoplanet, the accretion rates are significantly reduced and have cumulative mass profiles that are similar in shape.  This also indicates that the long-term accretion rates explored in the text at the beginning of Section 2.2.1 cannot be applied to massive planets in isolation. 

The point of this exercise is to demonstrate that planetesimal enrichment is dependent on the environment of the disk, not just the total mass in planetesimals and the average velocity dispersion.  In isolation, a protoplanet will quickly open a gap and starve itself from planetesimals.  However, a protoplanet formed by disk instability will be born in a disk with highly nonaxisymmetric structure, which will have an effect on the clump's accretion history. We use SIM2kmv2 in the next subsection to explore the role of a clump's formation environment on planetesimal capture.

\begin{figure}
\includegraphics[width=5.5cm, angle=-90]{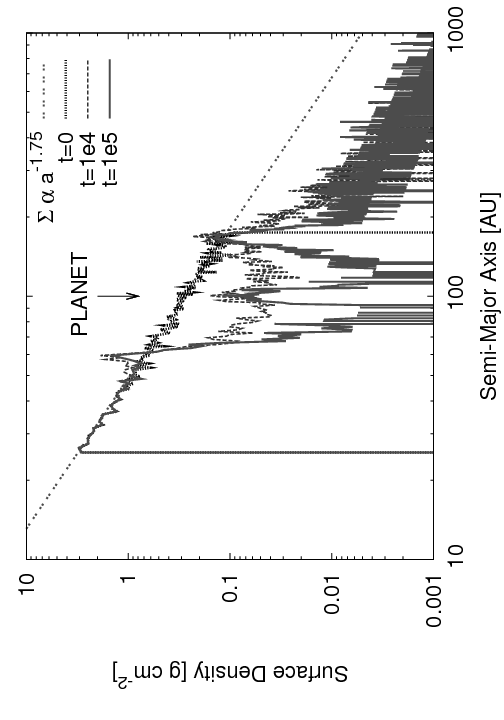}\includegraphics[width=5.5cm, angle=-90]{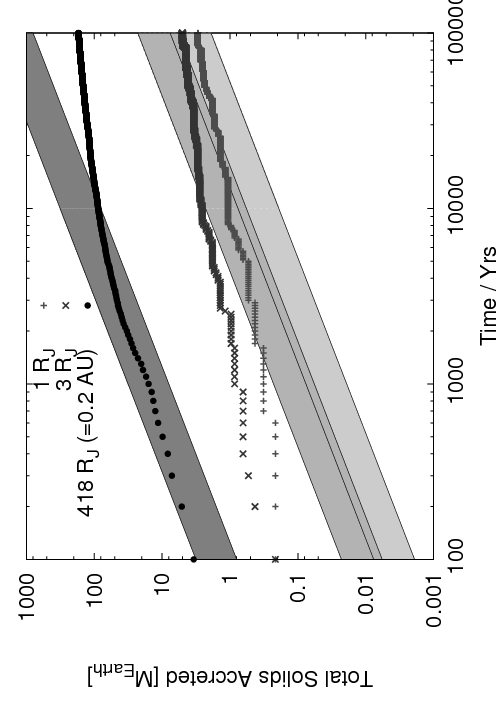}
\caption{Left: The surface density of planetesimals for three snapshots: 0, $10^4$, and $10^5$ yr. The initial surface density profile is set for a constant-$Q$ disk with Keplerian rotation and a temperature profile that falls as $a^{-0.5}$. Right: The total mass accreted by each protoplanet as a function of time is delineated by symbols, and the gray regions show the expected mass accretion for the velocity dispersion parameter $0.05<\alpha<0.1$. The accretion rates for the planets with small cross sections is subject to shotnoise effects, and are unreliable for roughly $t<10^4$ yr.  Without the presence of spiral arms, these protoplanets self-regulate their planetesimal accretion rates.}
\end{figure}


\subsubsection{Focusing Estimates and Velocity Dispersions from SIM2kmv2}

The environment in which disk instability and core assist gas giants form is dynamically different from the smooth disks envisaged in the scenario presented in section 2.2.2.  In a gravitationally unstable disk, planets and planetesimals are subject to strong perturbations from nonaxisymmetric structure, as well as from perturbations by other planets (see Fig.~1-4). Major accretion events occur when clumps collide and when they pass through spiral arms.  In the case of SIM2kmv2, the planetesimal flux through the clump is boosted during passages through planetesimal spiral arms.  In addition, the potential fluctuations of a spiral arm can compete with those of a proto-gas giant. This will tend to keep planetesimals in coherent structures that can cross the protoplanet's orbit (e.g., Fig.~5) and, perhaps, significantly delay gap opening.

 To explore the importance of environment, we return to the actual planetesimal encounter rate  measured in SIM2kmv2, for which 45 $M_{\oplus}$ material passed through the clump over 1230 yr.  This gives a measured $dM/dt\sim 0.04~\rm M_{\odot}~yr^{-1}$.   The average median mass of the clump is around 8 $M_J$ and has a volume-equivalent spherical radius of about 5 AU.  The measured average solid surface density at the radius of the clump is usually around 0.25 g cm$^{-2}$.  Equation (1) with the nominal values and with $a\sim110$ AU, which is consistent with the actual density, gives a $dM/dt\approx 0.004~\rm M_{\oplus}~yr^{-1}$ when the focusing term is ignored.  This suggests that the actual focusing in the simulation for this clump is about $F_g\sim 10$.   To determine whether the two-body focusing term used in equation (1) is reasonable, we need to know $\alpha$ from the simulation.  In Figure 8, we show four plots that give the total, radial, azimuthal, and vertical velocity dispersions for the planetesimals as a function of time and location in the disk.  The components of the dispersion are calculated using $\sigma_X=\left<v_X^2\right>-\left<v_X\right>^2$, where the averages (brackets) are calculated for all particles in an annulus centered on the location of interest and has a width that is two Hill radii for a 10 $M_J$ planet.   The dispersion begins around 0.1 $v_{\rm Keplerian}$, and grows to values between 0.2 and 0.3. The dispersion is still increasing at the end of the simulation, most notable near  the radii of the protoplanets (120 and 140 AU).  Gravitational instabilities first occur in the outer disk, which is where the dispersion first increases.  As the spiral arms develop, the excitation of planetesimals propagates to  smaller radii.    The increase in the vertical dispersion component before 500 yr is a result of the planetesimals being placed into the disk with too cold of an initial dispersion for the vertical extent over which they are distributed.  Although this may have an effect on the final dispersions we measure, the vertical dispersion alone does not account for the high total dispersion.  
Using these results and setting $M_{\rm planet}=8 M_J$, $a=110$ AU, and $R_{\rm cap}\approx 5$ AU, we find a gravitational enhancement factor to be about 10 for $\alpha=0.2$ and about 5 for $\alpha=0.3$, which seems to be in a fairly good agreement with the simulation. Whether this level of agreement continues through the clump's entire contraction sequence is unclear, and further investigation warrants a separate study.  Nonetheless, the current study provides an estimate for the velocity dispersions we should expect in a gravitationally unstable disk that forms fragments. It also suggests that spiral arms may play a fundamental role in the degree of enrichment a disk instability planet can achieve. \ACBc{We also remind the reader that, in the model envisaged here, violent bursts of disk instability due to, e.g., mass loading, represent transient, but critical phases of disk evolution and planet formation. We do expect quiescent  phases to occur between bursts (e.g., Vorobyov \& Basu 2006). }

\ACBc{The velocity dispersion of planetesimals that are decoupled from the gas during times of strong GI active is within the regime where destructive collisions are likely (Wyatt \& Dent 2002). From this result, one might worry that GIs would ultimately frustrate the growth of very large solids (see, e.g., Britsch et al.~2008).  However, a self-consistent model for planetesimal growth and destruction is required before the effects of high velocity dispersions on planetesimal evolution can be stated with certainty.  For example, if the planetesimals have a dust reservoir, then high velocities could enhance the growth of oligarchs (Xie et al.~2010).  Despite this uncertainty, it is worth estimating the collision rate between the km-size planetesimals to assess how quickly the planetesimal size distribution might change over the duration of our simulation.  Let the collision rate $C= 1.5 [\Sigma_z \alpha v_{\rm Keplerian} /(H R_p \rho_{\rm av} )] [1 + 16\pi G R_p^2 \rho_{\rm av}/(3(\alpha v_{\rm Keplerian})^2)]$, where $\rho_{\rm av}$ is the average internal density of the planetesimals and $R_p$ is the planetesimal size.  Assume $\rho_{\rm av}=3$ g cm$^{-3}$, $R_p=1$ km, $\Sigma_z=1$ g cm$^{-2}$, $v_{\rm Keplerian}=3.5$ km s$^{-1}$, $H=0.05a$, with $a=70$ AU, and $\alpha=0.2$. Using these numbers, the gravitational enhancement term is marginalized, and a given planetesimal should expect a collision rate $C\sim2\times10^{-7}$ per yr for the given semi-major axis.  In the event that the most of the solid mass is in km-size bodies, the evolution of planetesimal sizes will be slow compared with the duration of strong instability.} 

In sections 2.2.1, 2.2.2, and here, we have ignored the effects of gas accretion onto the proto-gas giant.  Any clump that survives to become a planet must evolve, for at least some time, in a massive disk.  Prodigious gas accretion could eventually erase any enrichment signature that the planet once had.  Moreover, one might worry whether the clump will eventually gain enough mass to evolve out of the planetary mass regime. We address these concerns in the following subsection.

\begin{figure}
\includegraphics[width=6cm, angle=-90]{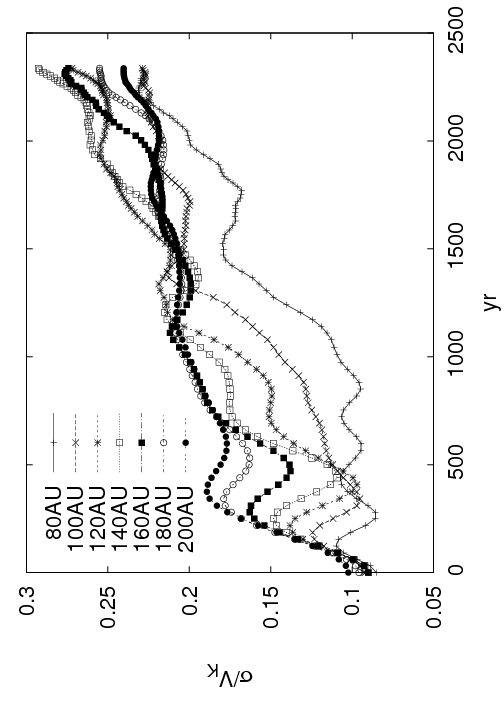}\includegraphics[width=6cm, angle=-90]{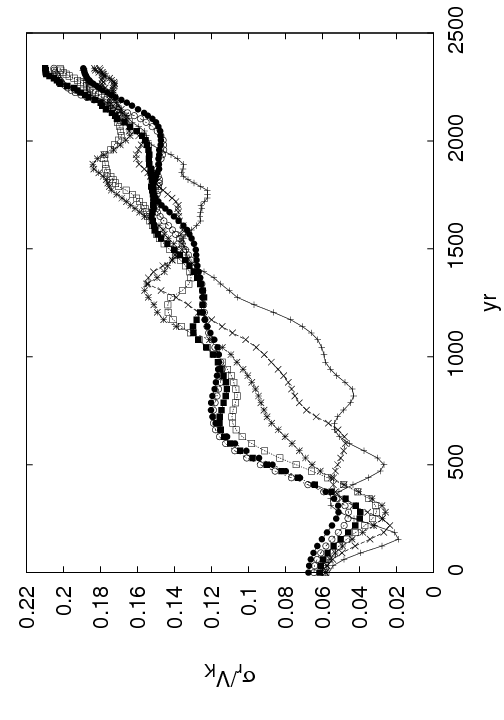}
\includegraphics[width=6cm, angle=-90]{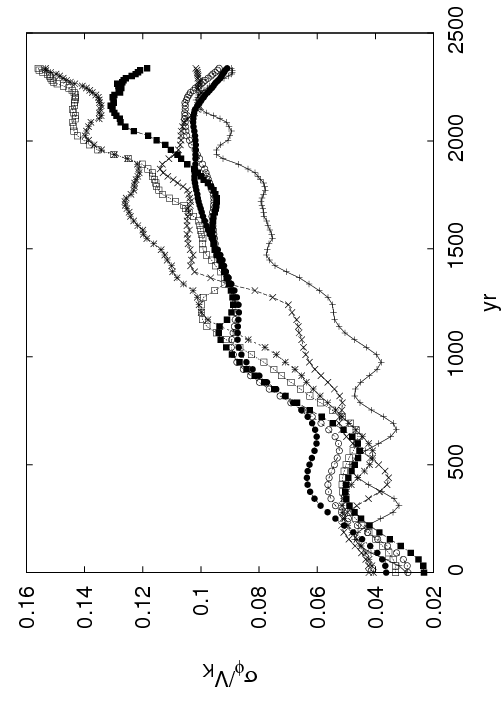}\includegraphics[width=6cm, angle=-90]{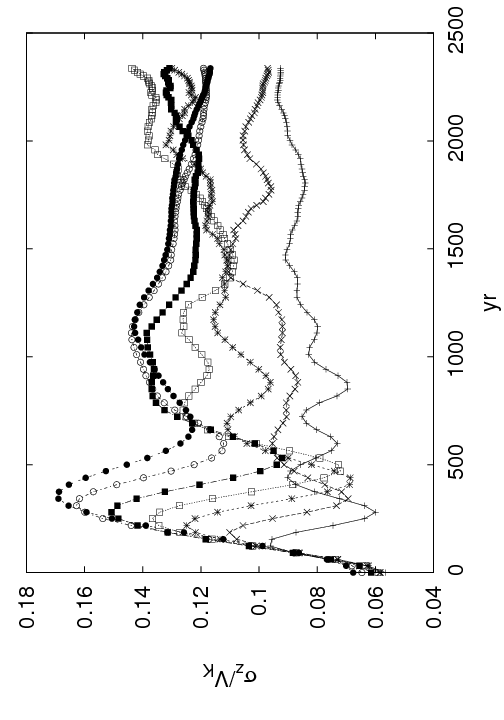}
\caption{Velocity dispersions for SIM2kmv2.  The dispersions are measured using averages from annuli that are two Hill radii for a 10 M$_J$ planet.  The different symbols correspond to the centers of the annuli.  The velocities are all shown relative to a Keplerian speed appropriate a solar-mass star.  }
\end{figure}


\subsubsection{Gas Accretion}

The above estimates do not include the possibility of gas accretion, which could dilute enrichment signatures.  Although this may happen in some cases, we do question whether appreciable gas accretion during the first-core phase or after dissociative collapse is a likely outcome for the following reasons.   First, any mass accretion during the first-core phase will only hasten the contraction time. For example, Boley et al.~(2010) found that, for clump masses  $\gtrsim8 M_J$, collapse occurred before the clump mass could be doubled.  Interestingly, collapse tended to occur around 10 $M_J$.  Second, these clumps can be on eccentric orbits, leading to phases of partial (or total) disruption. This would subject the clump to the competing effects of mass loss at pericenter and mass gain at apocenter.  Third, after the protoplanet undergoes dissociative collapse, its radius will be about 3 $R_J$.  These clumps will initially have considerable angular momentum, which will need to be mediated by a circumplanetary disk (e.g., Boley et al.~2010), potentially changing the timescale for mass growth for the planet.  Finally, suppose that gas mass accretion is inevitable, and that these clumps must eventually form brown dwarfs or low-mass stars (10 $M_J$ to 100 $M_J$; \ACBc{see arguments in Kratter et al.~2010}).   If the period during which disks typically remain massive lasts $\sim 10^5$ yr, e.g., the envelope infall timescale, the average accretion luminosity for such an object would then be $\sim G 10 M_J~100 M_J/(3R_J~{10^5~\rm yr})\sim 1 L_{\odot}$.  We cannot use such a rough estimate for the luminosity to conclude that  rapid growth cannot happen, but we can conclude that radiation from the growing clump will affect the behavior of the surrounding gas.   Regarding this point,  distinguishing between high-accretion-rate clumps in the outer disk and a bright central protostar would be a challenge with current instruments, but could be resolved using ALMA. 

\subsubsection{Summary of Enrichment By Planetesimal Capture}

Planetesimal capture can lead to a distribution of heavy element masses in disk instability planets, including gas giants on wide orbits that range from depleted to highly enhanced in heavy elements. The level of enrichment is sensitive to the velocity dispersion parameter $\alpha$,  the contraction time of the proto-planet, the size distribution of planetesimals, and planet-disk interactions.  For the latter, nonaxisymmetric structure in gravitationally unstable disks may lead to nontrivial differences in the final enrichments levels of planets, as a proto-planet in a smooth disk will starve itself from planetesimal accretion by opening a gap. 

\subsection{Differentiation Plus Tidal Disruption}

Instead of adding solids to a planet, heavy element-rich planets could be produced by removing a metal-poor envelope through tides.  During the early evolution of the first-core phase, disk instability planets are cold and extended, allowing solid material to settle to the planetary center even in the presence of convection (e.g., Decampli \& Cameron 1978; Boss 1997; Helled et al.~2008).  This differentiation occurs immediately in the clump after formation,
regardless whether the clump is born with well-mixed grains (Helled et
al. 2008) or whether large solids have been captured in spiral arms
prior to fragmentation (BD2010).  
An obvious outcome of such settling is core formation in disk instability planets (Helled \& Schubert 2008). However, differentiation may also be critical for the planet's final composition. When a clump forms, it
could be placed on an initially eccentric orbit at birth or can be
migrated/scattered inward by subsequent clump-disk and clump-clump
interactions (see Boley et al.~2010).  As the protoplanet's Hill sphere shrinks, mass can be
removed from the clump. Due to settling of solids, the outer gaseous envelope will typically be metal-poor, and the remaining planetary object will be metal-rich.  Tidal stripping (Boley et al.~2010; Nayakshin 2010a,b; Nayakshin 2011) will preferentially remove the metal-poor regions of the fragment. In the event that most of the envelope is removed by tidal stripping, the remaining core, if formed,  could eventually grow in mass in a way analogous to gas capture in the the core accretion scenario.  BD2010 call this situation core-assist plus gas capture, a mechanism that could lead to the formation of giant planets with very large cores.  


\section{Discussion of Enrichment of Disk Instability Planets} 

In the following subsections, we consider situations that could lead to both large and small high-Z fractions in disk instability gas giants.  We also discuss how we might expect the various enrichment mechanisms to trend with nebular metallicity.

\subsection{Gedanken Experiment with HAT-P-20b}

As mentioned in the introduction, the planet HAT-P-20b is estimated to consist of over 300 $M_{\oplus}$ of heavy elements (e.g., Leconte et
al.~2010). We do {\it not} argue here that this planet must have been formed by disk instability; instead, we simply argue that the planet's Brobdingnagian core is not proof of formation by core accretion.   In this subsection, we speculate on scenarios that could, in principle, lead to the formation of a planet like HAT-P-20b.  

(1) Assume that one or more fragments (clumps) form in the outer disk and
grow/coalesce to a mass of 15 $M_J$ before being scattered inward. Assume each fragment is enriched at birth near the maximum value of about 2  (see section 2.1).   Altogether, a 15 $M_J$ clump with a total enrichment at birth of $\sim$
2 and metallicity 2.2 over Solar (HAT-P-20) will have about 420
$M_{\oplus}$ of heavy elements.  
 As the protoplanet moves inward in radius,
its Hill sphere shrinks enough to remove a large fraction of the gaseous envelope, which is preferentially metal-poor due to settling.  This tidal stripping removes gas until the planet becomes the 7 $M_J$ planetary object with the large core seen today. Complete tidal disruption of clumps is seen in several simulations, including Figure 1 in this work
(see also Boley et al. 2010; Vorobyov \& Basu 2010; Cha \& Nayakshin 2010), so removing
half of the planet's mass through tides is not unreasonable. 

(2) Assume that the fragment forms and grows to about 6 $M_J$ during the first-core phase.  The clump is transported inward, but not fast enough to lead to overall mass loss.  If the velocity dispersion of comet-like planetesimals at 10s of AU has only been pumped to 0.1 times the Keplerian speed by the GIs at 100 AU, an appreciable amount of solids can be captured during each pericenter passage.  The clump attains $\sim 300$ $M_{\oplus}$ of solids from planetesimal capture alone, giving a total mass of  about 7 $M_J$ for the planet.

(3) A combination of (1) and (2) occurs. 

(4) The clump from (1) or (3) is completely destroyed by tides at pericenter.  The large core of heavy elements remains bound, and begins a second phase of gas accretion. In this scenario, HAT-P-20b would represent a core assist plus gas capture planet.

If a disk instability planet is to become a HAT-P-20b-like planet, considerable inward migration is necessary during any of the above scenarios, which is also required in a core accretion scenario. 

\subsection{Metal-Poor Gas Giants}

SIM2kmv2 shows that if many solids are trapped in large bodies, a fragment can be born with subnebular metallicity.  Strictly, we expect the km-size planetesimals to be captured eventually as the clump contracts to higher densities, an effect not resolved in the global simulations presented here.  However, if most of the mass of heavy elements is in very large planetesimals, e.g., 10s of kilometers or larger, clumps may be unable to capture aerodynamically enough high-Z material to bring the metallicity back to nebular values.  For example, using the $R_{\rm cap}$ radii from HB2010 for 100 km planetesimals and $\alpha=0.3$, the fraction of captured planetesimals, under the same assumptions used in Section 3, is about 0.4 times the well-mixed fraction for a 7 $M_J$ clump.  This effect is quite sensitive to the planetesimal size and velocity distributions.  When $\alpha=0.1$, even 100 km-size planetesimals can be captured with an overall $f_{\rm enr}\sim3$ for a 7 $M_J$ clump.  A combination of rapid contraction, high velocity dispersion, and large planetesimals may be required to keep a clump from returning to nebular composition. 

\subsection{Trends for Enrichment and Planet Frequency vs.~Host Star Metallicity}

From the above discussion, it should be clear that disk instability planets can have a wide variety of enrichments, from metal-poor to metal-rich.  Frustratingly, core accretion, disk instability, and hybrid formation scenarios will likely lead to planets with a large overlap in compositions.  Distinguishing formation modes may only be possible by looking at enrichment trends.  

{\it Enrichment at Birth:}  If the capture efficiency of solids in spiral arms remains about the same for a wide range of metallicity, which is suggested by the simulations presented here, then we would expect a flat enrichment trend compared with planets formed purely from well-mixed gas of nebular composition.  Variations in planetesimal distributions will lead to scatter in the trend, unless the distribution of planetesimals sizes is strongly dependent on the metallicity of the nebula.  Such behavior may appear as a strong break, where, e.g., the streaming instability becomes highly efficient (Johansen et al.~2009).

{\it Core Assist Plus Gas Capture:} As the opacity of a system increases, the frequency of {{\it transient}} clumps should also increase (BD2010).  In this case, we would expect the frequency of gas giants to increase with metallicity.  This would be consistent with expectations from core accretion plus gas capture, but the two mechanisms may not have the same scaling nor may they have the same degree of enrichment on average.  Further work is required to explore differences between these formation channels.

{\it Planetesimal Capture:}  Based on Table 3, this mechanism is the most promising for producing a large degree of heavy element enrichment.  From equation (1), we expect the mass of solids to scale with the metallicity of the disk (see also Helled \& Schubert 2009).  From this alone, there should be no gas giant enrichment trend with metallicity when compared with a well-mixed composition.  However,  as the number of planetesimals in the disk increases, the effects of the accretion energy deposition will become more pronounced.  This will tend to delay the protoplanet's dissociative collapse time.    Because the contraction time is the principal determining factor in the degree of heavy element enrichment by planetesimals, the fraction of solids that will be captured will increase.  If the dust from planetesimal ablation can replenish dust settling above the photosphere, the effect will be further exacerbated.   Overall, the enrichment of disk instability planets will trend strongly with host star metallicity if planetesimals are present at the time of fragmentation.  As with enrichment at birth, there may be a break in the trend, directly related to the distribution of planetesimals as a function of nebular metallicity.

Both enrichment at birth and planetesimal capture are dependent on the formation of large solids shortly before the disk fragments.  If large solids have not formed at this time, grain sedimentation followed by tidal stripping  seems to be the only way disk instability can produce extremely heavy element enriched gas giants.  Our understanding of gas giant planet formation, whether in the context of disk instability or core accretion, is dependent on a broad understanding of the formation of large solids. 

\section{Summary}

We show that disk instability gas giants can be born with a variety of heavy element enrichments, ranging from sub- to super-nebular metallicities. In addition, planetesimal capture during the pre-collapse (first core) phase of evolution could further enrich the protoplanet, possibly creating gas giants with large heavy element masses. Differentiated clumps that under go partial tidal stripping during their first core phase could also lead to a planet that is enriched in heavy elements.  For the case of planetesimal accretion, nonaxisymmetric structure due to GIs may prove to be critical in delaying proto-gas giants from opening a planetesimal gap and starving themselves from heavy element enrichment. Populations of planetesimals and dwarf planets that have large pericenters and form a scattered disk may be relics of a period of outerdisk clump formation.

In light of the multiple pathways available for enriching disk instability gas giants, individual planets with large heavy element masses (e.g., HAT-P-20b), $M_{\rm core}\sim 300 M_{\oplus}$ cannot be taken as {\it prima facie} evidence for the planet's formation mechanism, in agreement with Helled \& Schubert (2009).   Moreover, it is not obvious that standard core accretion provides a route for creating heavily enriched, massive gas giant planets, as the core mass required to initiate rapid gas capture can be small when compared with the well-mixed distribution of solids.  Additional theoretical and observational studies are required to better
understand whether population trends as a function of, e.g., host star metallicity can allow us to determine the relative importance of  different formation mechanisms.  We note that if the dominant enrichment mechanism for disk instability gas giant planets is planetesimal accretion, the degree of enrichment, relative to a well-mixed composition, should trend with host star metallicity. The degree of this trend has yet to be determined.  

{We thank P.~Bodenheimer,  L.~Mayer, \ACBc{Eric Ford, and the anonymous referee} for comments that improved this manuscript.} ACB's support was provided by a contract with the California Institute of Technology (Caltech) funded by NASA through the Sagan Fellowship Program. R.H. acknowledges support from NASA through the Southwest Research Institute.  M.J.P was supported by NASA Origins of Solar Systems grant NNX09AB35G. Resources supporting this work were provided by the NASA High-End Computing (HEC) Program through the NASA Advanced Supercomputing (NAS) Division at Ames Research Center.

\end{document}